\documentclass[]{raa}            % referee version: for submission
\usepackage{graphicx,times}
\usepackage{natbib}

\providecommand{\bm}[1]{\mbox{\boldmath$#1$\unboldmath}}

\begin{document}

%\title{Indication of 1860 Hz quasi-periodic oscillations from a neutron star 
%low-mass X-ray binary
%}
%\title{Plausible 1860 Hz quasi-periodic oscillations simultaneously observed with a pair of 
%kilohertz quasi-periodic oscillations
%}
\title{Simultaneous observations of a pair of kilohertz QPOs and a plausible 1860 Hz QPO
from an accreting neutron star system
}

\volnopage{ {\bf 2009} Vol.\ {\bf X} No. {\bf XX}, 000--000}
   \setcounter{page}{1}

\author{Sudip Bhattacharyya
\inst{}
}

\institute{Department of Astronomy and Astrophysics, Tata Institute
of Fundamental Research, Mumbai 400005, India; {\it sudip@tifr.res.in}\\
\vs \no
{\small Received [year] [month] [day]; accepted [year] [month] [day] }
}

\abstract{
We report an indication  ($3.22 \sigma$) of $\approx 1860$ Hz quasi-periodic oscillations 
from a neutron star low-mass X-ray binary 4U 1636-536. If confirmed,
this will be by far the highest frequency feature observed from an accreting neutron star system,
and hence could be very useful to understand such systems. This plausible timing feature
was observed simultaneously with lower ($\approx 585$ Hz) 
and upper ($\approx 904$ Hz) kilohertz quasi-periodic oscillations.
%and its frequency was close to the triple and the double of the frequencies of the latter two respectively.
The two kilohertz quasi-periodic oscillation frequencies had the ratio of 
$\approx 1.5$, and the frequency of the alleged $\approx 1860$ Hz feature
was close to the triple and the double of these frequencies. This can be
useful to constrain the models of all the three features. In particular,
the $\approx 1860$ Hz feature could be (1) from a new and 
heretofore unknown class of quasi-periodic oscillations, or (2) the first observed overtone 
of lower or upper kilohertz quasi-periodic oscillations.
Finally we note that, although the relatively low significance of the $\approx 1860$ Hz 
feature argues for caution, even a $3.22 \sigma$ feature at such a uniquely high frequency
should be interesting enough to spur a systematic search in the archival data, as well as
to scientifically motivate sufficiently large timing instruments 
for the next generation X-ray missions.
}

\authorrunning{Sudip Bhattacharyya}
\titlerunning{1860 Hz quasi-periodic oscillations}
\maketitle

\section{Introduction}\label{Introduction}

Quasi-periodic oscillations (QPOs) are observed from many neutron star low-mass
X-ray binary (LMXB) systems \citep{vanderKlis2006}. There are several kinds of QPOs,
such as millihertz QPO, horizontal branch QPO, normal/flaring branch QPO, hectohertz
QPO, lower kilohertz (kHz) QPO, upper kHz QPO, etc. \citep{vanderKlis2006}. 
In this paper we will mostly concentrate on high frequency QPOs, that can 
currently be observed only with the proportional counter array (PCA) instrument
of the {\it Rossi X-ray Timing Explorer} ({\it RXTE}) space mission. KHz QPOs
are such high frequency QPOs, and often they appear as a pair. For a given source,
their frequencies normally move up and down together in the $\approx 200-1200$ Hz range
in correlation with the source state. 
%The separation between lower and upper
%kHz QPO frequencies is typically within 20\% of the neutron star spin frequency
%or half of that depending on the source \citep{vanderKlis2006, MendezBelloni2007}. 
KHz QPOs are scientifically very important for the following reasons:
(1) their high frequencies indicate that
they originate from regions close to the neutron stars, and hence could be
used as a tool to probe the strong gravity region around these stars, as well
as the neutron star properties, and
(2) they have been observed from many sources, and repeatedly from a given source,
with high significance.
Indeed, there are several models that involve
relativistic orbital frequencies and the neutron star spin frequency, as well as 
beating and resonances among them \citep{StellaVietri1998, Lamb+1985,
Miller+1998, AbramowiczKluzniak2001, Torok+2008, Wijnands+2003, LambMiller2003,
Zhang2004, Mukhopadhyay2009}.
However, although there are many kHz QPO models available in the literature,
none of them can explain all the major properties of this timing feature. 
Therefore, these QPOs cannot yet be 
used as a reliable tool to probe the strong gravity region, or to measure
the neutron star parameters.

An intriguing aspect of the kHz QPOs is that so far they have never been 
observed with a frequency greater than 1330 Hz \citep{vanderKlis2006}, while PCA can easily detect
a feature with a much higher frequency. In fact, even this 1330 Hz QPO
was not confirmed by a later analysis \citep{Boutelieretal2009b}.
This implies that no timing feature has ever been observed with $> 1330$ Hz
from an accreting neutron star system.
In this paper, we report the indication
of an $\approx 1860$ Hz QPO from the persistent neutron star LMXB 4U 1636-536 
(see \citet{Altamirano+08} for the other timing features).
This plausible QPO was observed simultaneously with the pair of kHz QPOs,
which suggests that the $\approx 1860$ Hz QPO could be from a new and previously unknown
class of very high frequency QPOs. Note that kHz QPOs from 4U 1636-536 were first
discovered by \citet{vanderKlisetal1996, Zhangetal1996a, Zhangetal1996b}, 
and later were reported by many authors \citep{Wijnandsetal1997, Vaughanetal1997,
Zhangetal1997, Vaughanetal1998, Psaltisetal1998, Mendezetal1998, 
Mendez1998, MisraShanthi2004, Kaaret+1999, Psaltisetal1999, 
Markwardtetal1999, Jonkeretal2000, Fordetal2000,
Mendez2002, Jonkeretal2002, DiSalvoetal2003, MisraShanthi2004, Jonkeretal2005, 
Barretetal2007, Bellonietal2007, Torok+2008}.

\section{Data Analysis and Results}\label{DataAnalysisandResults}

The LMXB 4U 1636-536 was observed with {\it RXTE} PCA on September 28, 2007. We
analysed the data from the event file FS4f\_19d86be0-19d8c749 
(time: 16:17:36 to 22:47:38, i.e., 23.402 ks duration) 
of the obsID 93091-01-01-000, and searched for QPOs. We excluded the
data gaps and a burst, as well as the time intervals in which the 
number of used proportional counter units
(PCU) was changed, and divided the 21.85 ks of data into $M$ (= 2185) equal segments of 10 s
duration. We performed a fast Fourier transform on each time segment for the entire
PCA energy range, and 
the resulting power spectra (up to 2048 Hz) were averaged in order to reduce the noise.
By combining $W$ (= 4) consecutive frequency bins (making the frequency resolution 
0.4 Hz), we found a tentative peak at the frequency $1861.45$ Hz (see the upper panel of
Fig.~1). Before exploring this peak further, we binned the power spectrum more
in order to search for kHz QPOs.
This gave the clear evidence of two kHz QPOs at $\approx 585$ Hz and $\approx 904$ Hz (see the lower
panel of Fig.~1). For probing the properties of these QPOs, we fitted the power 
spectrum with a model, and minimized the corresponding $\chi^2$ to
get the best-fit parameter values.
This model was the sum of a ``constant" (to account for the Poisson noise level including 
the effect of deadtime), a ``powerlaw" (to describe the red noise or broad-band feature
at lower frequencies), and two Gaussians (to fit the kHz QPO pair). 
Then we divided the power spectrum with the continuum part of the
best-fit model (see, for example, the lower panel of Fig.~1) 
and multiplied by 2, in order to have the noise distributed as 
$\chi^2$ with $2MW$ degrees of freedom \citep{vanderKlis1989, Bhattacharyya+06}. 
The peak powers of the 585 Hz and 904 Hz QPOs were 2.00669 (for $W=1024$) and 2.00959 (for $W=2048$)
respectively, which gave the respective single trial significance of $2.87\times10^{-7}$
and $2.20\times10^{-24}$. Therefore, multiplying by the number of trials (20 and 10),
we got a significance of $5.74\times10^{-6}$ (i.e., $4.5\sigma$) and $2.20\times10^{-23}$
respectively. These numbers show that both the kHz QPOs were very significant.
The Gaussian modeling gave the RMS amplitudes $\approx 4.5$\% and $\approx 7$\%, and
the quality factors $\approx 5.6$ and $\approx 7.3$ for the lower and the upper kHz QPOs
respectively.

In order to estimate the significance of the observed peak at $\approx 1860$ Hz, we followed
the same procedure as mentioned in the previous paragraph. After having the noise 
distributed as $\chi^2$ with $2MW$ degrees of freedom, the peak power 2.1094 gave
the single trial significance of $2.47\times10^{-7}$. 
We considered the number of trials to be equal to the number of frequency bins ($= 5120$)
in the power spectrum. This is because, (1) we applied no cuts to the data (e.g., based
on energy bands), apart from the standard filtering for burst and gaps; and
(2) we did not search for QPOs in any other data set of 4U 1636-536.
Note that we chose this particular data set in connection with a research on broad iron lines,
and we happened to find the feature at $\approx 1860$ Hz. Furthermore, since the 
frequency of high frequency QPOs evolves with time \citep{vanderKlis2006}, it is unlikely
that a feature with the {\it same} frequency will appear in another observation.
Therefore, it should be reasonable to consider only the current data set in order to
estimate the number of trials. Hence, considering the 
number of trials $= 5120$, we found that the plausible QPO
at $\approx 1860$ Hz was $3.22 \sigma$ significant. The RMS amplitude
of this feature was $\approx 1\%$, and it had an extremely high quality factor ($Q \approx
4650$), as this QPO was observed with 0.4 Hz resolution.

\begin{figure}[h]
\centering
\includegraphics[width=9.0cm, angle=0]{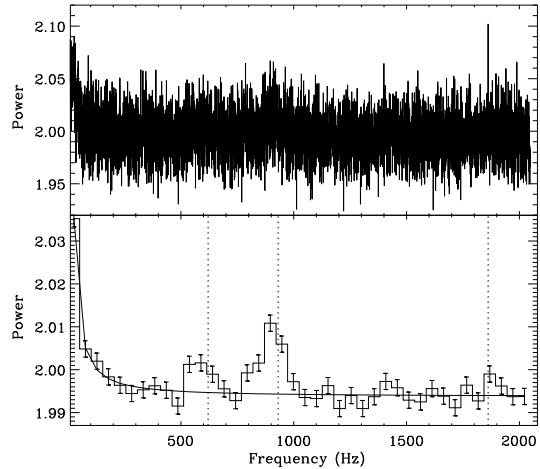}
\begin{minipage}[]{85mm}
\caption{Power spectrum (averaged over 2185 spectra; each from a 10 s time segment) 
of the 21.85 ks {\it RXTE} PCA data from the source 4U 1636-536 (see text). 
Upper panel (power spectrum with 0.4 Hz frequency resolution): 
a peak near 1860 Hz is clearly seen.
Lower panel (same power spectrum 
with 51.2 Hz frequency resolution): two kHz QPOs (near 585 Hz and 
904 Hz) are clearly seen. $1 \sigma$ error bars and a model (solid line; describing the continuum
part of the power spectrum) are shown. The vertical dotted lines mark the frequency of the
narrow $\approx 1860$ Hz peak of the upper panel, and its half and one-third
frequencies. Note that the latter two pass through the upper and lower kHz QPOs respectively.} \end{minipage}
\label{specfirst}
\end{figure}

\begin{figure}[h]
\centering
\includegraphics[width=9.0cm, angle=0]{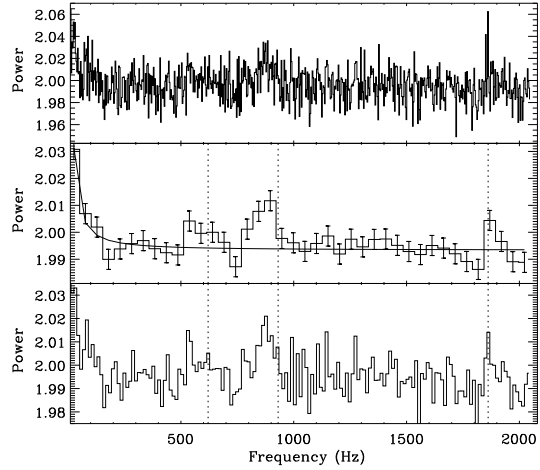}
\begin{minipage}[]{85mm}
\caption{Power spectrum (averaged over 560 spectra; each from a 10 s time segment)
of the 5.6 ks {\it RXTE} PCA data from the source 4U 1636-536 
(see \S~\ref{DataAnalysisandResults}). Upper panel 
(power spectrum with 3.2 Hz frequency resolution): a peak near 1860 Hz is clearly seen.
Middle panel (same power spectrum with 51.2 Hz frequency resolution): two kHz QPOs (at the
same frequencies shown in Fig.~1) are clearly seen. A tentative broad peak near 1860 Hz 
is also seen.
$1 \sigma$ error bars and a model (solid line; describing the continuum
part of the power spectrum) are shown. The vertical dotted lines mark the frequency of the
narrow $\approx 1860$ Hz peak of the upper panel, and its half and one-third
frequencies. Note that the latter two pass through the upper and lower kHz QPOs respectively.
Lower panel (same power spectrum with 12.8 Hz frequency resolution): the narrow part
(marked with a vertical dotted line) and the broad part (suggested from the powers of
several adjacent frequency bins above the average power) of the plausible $\approx 1860$
Hz QPO are visible.} \end{minipage}
\label{powspec2}
\end{figure}

In order to find out whether the plausible $\approx 1860$ Hz QPO was stronger
in a part of the dataset, we examined the power spectra of nine unequal data segments.
This division was naturally done in order to exclude the data gaps, burst, etc. (see
the first paragraph of \S~\ref{DataAnalysisandResults}).
%and without any prior knowledge of the variation of QPO strength. 
We searched for a plausible timing feature at the known frequency
of $\approx 1860$ Hz in the power spectrum of each of these data segments.
Although we found the indication of such a feature 
in several segments, it was strong in the power spectrum of only one segment 
(hereafter segment 1; time: 19:20:06 to 20:53:26, i.e., 5.6 ks duration). This power spectrum was calculated
by averaging the spectra of $M$ (= 560) equal segments of 10 s duration. By
combining $W$ (= 32) consecutive frequency bins (making the frequency resolution
3.2 Hz), we found a peak at the frequency $1860.85$ Hz (see the upper panel of
Fig.~\ref{powspec2}). After having the noise distributed as $\chi^2$ with $2MW$ degrees of 
freedom (see the first paragraph of \S~\ref{DataAnalysisandResults}), we got a peak power of 2.07.
Therefore, for the search at a known frequency in this particular power spectrum, 
the $\approx 1860$ Hz QPO had the significance of $1.82\times10^{-6}$ 
(i.e., $4.8 \sigma$). The RMS amplitude of this QPO was $2.4\%$.
%and the quality factor was $\approx 580$. 
Therefore, this QPO was stronger in segment 1
than in the whole data set. Here we note that 
even for the segment 1, the $\approx 1860$ Hz QPO occured simultaneously with
the lower and the upper kHz QPOs (see the middle panel of Fig.~\ref{powspec2}).

The very high $Q$-value of the plausible $\approx 1860$ Hz QPO is somewhat
unusual. We, therefore, searched for a broader feature 
at the known 1860 Hz frequency in the power spectrum of segment 1, and indeed 
we found a tentative peak with the centroid frequency of 1868.85 Hz (middle panel of Fig.~\ref{powspec2}).
After having the noise distributed as $\chi^2$ with $2MW$ degrees of freedom,
the peak power 2.0109 gave a significance of $1.82\times10^{-3}$ (i.e., $3.1\sigma$).
This tentative broad feature had the RMS amplitude of $\approx 3.7\%$, and a more usual
quality factor of $\approx 35$. Note that, although such a broad feature is not 
significant in the power spectrum of the whole dataset, the lower panel of Fig.~1
does show a small broad peak at the frequency of the plausible narrow $\approx 1860$ Hz QPO.
Therefore, while the evidence of a broad feature is weak, we still explored the data a little
more in order to see whether the plausible narrow QPO could be a part of a broad QPO. The
upper and the middle panels of Fig.~\ref{powspec2} exclusively show the plausible narrow QPO and the
plausible broad QPO respectively. Therefore, in the lower panel of the same
figure, we used an intermediate frequency resolution of 12.8 Hz, and found
both the plausible narrow QPO (marked with a dotted vertical line) and 
the plausible broad QPO (suggested from the powers of several adjacent bins above
the average power). This indicates that the narrow QPO at $\approx 1860$ Hz could be
a part of a broader QPO.

Finally, the vertical dotted lines of the figures show that the frequency of 
the plausible $\approx 1860$ Hz QPO is close to the double of the upper kHz QPO
frequency and triple of the lower kHz QPO frequency. Therefore, the $\approx 1860$ Hz QPO
could be the first overtone of the upper kHz QPO or the second overtone of the lower 
kHz QPO. 

\section{Discussion}\label{Discussion}

In this paper, we report the detection of a pair of kHz QPOs from a neutron star
LMXB 4U 1636-536. The ratio of these QPO frequencies is roughly 1.5, which
is somewhat consistent with the resonance model described in 
\citet{Abramowiczetal2003} (but see \citet{Boutelieretal2009a}; see also
\citet{Zhangetal2006}). It is generally observed that the separation of the 
twin kHz QPO frequencies clusters around the neutron star spin frequency,
or half of that. Models have been proposed in order to explain this aspect.
However, for the kHz QPOs reported in this paper, the separation 
is $\approx 319$ Hz, which is quite different from the half of the stellar
spin frequency (see also \citet{MendezBelloni2007, Yinetal2007}). Note that the 
neutron star of 4U 1636-536 spins with a rate of 582 Hz \citep{StrohmayerMarkwardt2002}.

Apart from the  kHz QPOs, we have found an indication of a QPO at $\approx 1860$ Hz.
This is by far the highest frequency feature observed from an accreting neutron star system,
and hence could be extremely useful to understand such systems. Moreover, this plausible
QPO was observed simultaneously with the lower and the upper kHz QPOs, which means that the
$\approx 1860$ Hz QPO could be from a new class of QPOs.

\begin{figure}[h]
\centering
\includegraphics[width=9.0cm, angle=0]{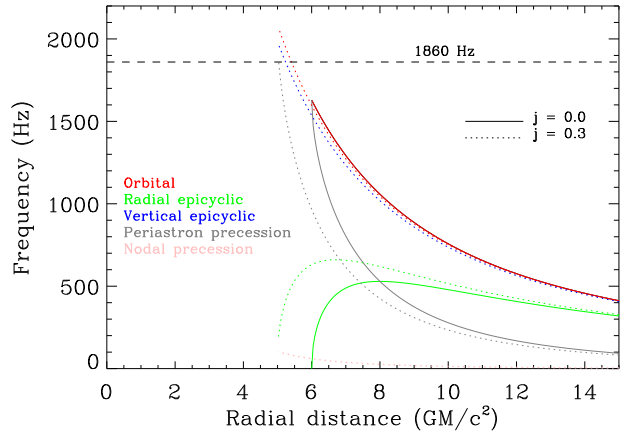}
\begin{minipage}[]{85mm}
\caption{Radial profiles of various frequencies (colour coded) of equatorial circular 
orbits in Kerr spacetime (see \S~\ref{Discussion}). 
Two angular momentum parameters ($j = 0.0$ (solid) and $j = 0.3$ (dotted)),
and neutron star mass $= 1.35 M_\odot$ are used. Dashed horizontal line exhibits the 
1860 Hz frequency. This figure shows that these frequencies, that are often used to explain
kHz and some other QPOs, cannot possibly explain a $\approx 1860$ Hz QPO.
}\end{minipage}
\label{freq1}
\end{figure}

We will now briefly discuss the plausible origin of this QPO using general 
arguments, and without going into specifics. 
In some of the models, the lower and upper kHz QPO and some other QPO frequencies are thought to be
one or more of the following frequencies (for equatorial circular orbits 
in Kerr spacetime; \citet{vanderKlis2006}): (1) Keplerian orbital frequency 
$\nu_{\phi} = \nu_{K}(1+j(r_g/r)^{3/2})^{-1}
= (2\pi)^{-1}(GM/r^3)^{1/2}(1+j(r_g/r)^{3/2})^{-1}$; (2) radial 
epicyclic frequency $\nu_r = \nu_{\phi}(1-6(r_g/r)+8j(r_g/r)^{3/2}-3j^2(r_g/r)^2)^{1/2}$; 
(3) vertical epicyclic frequency $\nu_{\theta} = 
\nu_{\phi}(1-4j(r_g/r)^{3/2}+3j^2(r_g/r)^2)^{1/2}$;
(4) periastron precession frequency $\nu_{\rm peri} = \nu_{\phi} - \nu_r$; and
(5) nodal precession frequency $\nu_{\rm nodal} = \nu_{\phi} - \nu_{\theta}$.
Here the last four frequencies are for
infinitesimally tilted and eccentric orbits, $M$ is the neutron star mass, $r_g =
GM/c^2$, $j = Jc/GM^2$, $J$ is the total angular momentum of the neutron star, and $r$
is the distance of an orbit from the centre of the neutron star.
For a sample of radio pulsars, $M$ was found to be 
distributed around $1.35 M_\odot$ in a narrow Gaussian ($\sigma = 0.04 M_\odot$;
\citet{ThorsettChakrabarty1999}). Therefore, considering $M = 1.35 M_\odot$,
we find that only $\nu_{\phi}$ or $\nu_{\theta}$ can be the frequency
of the plausible $\approx 1860$ Hz QPO, and that too for relatively large angular
momentum parameter values (as far as a neutron star is concerned) and for orbits very
close to the innermost stable circular orbit (ISCO; see Fig.~\ref{freq1}). As the neutron 
star mass increases due to accretion,
$M$ is expected to be greater than $1.35 M_\odot$ for
accreting neutron stars (such as the one in 4U 1636-536). Moreover, since
$\nu_{K} = (c^3/2\pi G)(r/r_g)^{-3/2}(1/M)$, $\nu_{K}$ decreases as $M$ increases
for a given $r/r_g$. Therefore, it is unlikely that any of the above mentioned
five frequencies can be the frequency of the plausible $\approx 1860$ Hz QPO. Besides,
since the spin frequency of the neutron star in 4U 1636-536 is $\nu_{\rm spin}
= 582$ Hz, the beating of $\nu_{\rm spin}$ with 
any of $\nu_{\phi}$, $\nu_r$ or $\nu_{\theta}$
cannot explain the frequency of this QPO. However, the $\approx 1860$ Hz QPO
could be an overtone of any of $\nu_{\phi}$, $\nu_r$, $\nu_{\theta}$ or
$\nu_{\rm peri}$, or beating between an overtone and a fundamental of these frequencies.
This plausible QPO could also be an overtone of one of the kHz QPOs 
(see \S~\ref{DataAnalysisandResults}).

Although it is unlikely that the $\approx 1860$ Hz QPO frequency is 
a Keplerian frequency $\nu_{\phi}$, it is still instructive to examine
what constraints $\nu_{\phi} = 1860$ Hz can impose on the neutron
star parameter values. This is because, the measurement
of neutron star parameters provides the only way to constrain
the theoretically proposed equation of state (EoS) models of neutron star
cores, and hence to understand the nature of supranuclear core matter
\citep{LattimerPrakash2007, Bhattacharyyaetal2000, Bhattacharyyaetal2001a}.
Various authors suggested ways to constrain the neutron star parameters assuming
one of the kHz QPO frequencies as a Keplerian frequency \citep{Kaaretetal1997,
Miller+1998, Zhangetal2007, Zhang2009}. For example, this assumption, and 
the following two reasonable conditions can constrain the mass ($M$) and the
radius ($R$) of a neutron star \citep{Miller+1998}:
\begin{eqnarray}
R \le r,
\label{freq4}
\end{eqnarray}
where $r$ is the radius of the orbit associated with
the kHz QPO via the expression of $\nu_{\phi}$; and
\begin{eqnarray}
r_{\rm ISCO} \le r,
\label{freq5}
\end{eqnarray}
where $r_{\rm ISCO}$ is the radius of the ISCO.
This is because the first condition gives a mass-dependent upper
limit on $R$ via the expression of $\nu_{\phi}$; and the second condition
gives an upper limit on $M$: $M < c^3/(2\pi 6^{3/2}G\nu_\phi|_r)$
(for Schwarzschild spacetime). Therefore, if 1860 Hz were a Keplerian frequency,
then the upper limit of neutron star mass and radius would be $\approx 1.2 M_\odot$
and $\approx 10.5$ km respectively. Such upper limits would support the strange
star EoS models \citep{Chengetal1998, Bhattacharyyaetal2001b}, although the less 
exotic neutron star EoS models could not be completely ruled out.

The uniquely high frequency of the plausible $\approx 1860$ Hz QPO and its 
coexistence with both the kHz QPOs make it extremely interesting. 
Furthermore, although no such high frequency signal was claimed previously,
\citet{vanderKlis2006} mentioned, ``the distinct impression of the observers is that 
there is still much hiding below the formal detection levels".
It should, therefore, be worthwhile to report
this feature, which will spur a systematic search of such high frequency features
in the archival {\it RXTE} PCA data, as well as in the large area xenon proportional 
counters (LAXPC) instrument data of the upcoming {\it Astrosat} space mission.
Moreover, next generation X-ray timing instruments, such as the high timing resolution 
spectrometer (HTRS; proposed for the {\it International X-ray Observatory})
and the proposed Si pixel detector of the {\it Advanced X-ray Timing Array}, 
will have much better capability to detect weak QPOs
(see, for example, Fig.~2 of \citet{Barret+08}; \citet{Chakrabarty+08}). 
The plausible $\approx 1860$ Hz QPO
will therefore motivate the future X-ray timing instruments.

\section*{Acknowledgments}

The paper was improved by the suggestions of the referee.

\label{lastpage}

\end{document}